\documentclass[twocolumn,english,british,aps,prl,superscriptaddress,groupedaddress]{revtex4} 
\usepackage[LGR,T1]{fontenc}
\usepackage{textcomp}
\usepackage{amstext}
\usepackage{graphicx}

\usepackage{color,soul}
\usepackage{hyperref}
\usepackage[sort&compress]{natbib}
\makeatletter

\DeclareFontEncoding{LGR}{}{}
\DeclareTextSymbol{\~}{LGR}{126}

\@ifundefined{textcolor}{}
{%
 \definecolor{BLACK}{gray}{0}
 \definecolor{WHITE}{gray}{1}
 \definecolor{RED}{rgb}{1,0,0}
 \definecolor{GREEN}{rgb}{0,1,0}
 \definecolor{BLUE}{rgb}{0,0,1}
 \definecolor{CYAN}{cmyk}{1,0,0,0}
 \definecolor{MAGENTA}{cmyk}{0,1,0,0}
 \definecolor{YELLOW}{cmyk}{0,0,1,0}
}

\makeatother

\usepackage{babel}
\begin{document}

\title{Active Temporal Multiplexing of Photons

}

\author{Gabriel J. Mendoza, Raffaele Santagati, Jack Munns, Elizabeth Hemsley,
Mateusz Piekarek, Enrique Mart\'{i}n-L\'{o}pez, Graham D. Marshall, Damien
Bonneau, Mark G. Thompson, and Jeremy L. O\textquoteright Brien}

\selectlanguage{british}%

\affiliation{Centre for Quantum Photonics, H. H. Wills Physics Laboratory \& Department
of Electrical and Electronic Engineering, University of Bristol, Merchant
Venturers Building, Woodland Road, Bristol, BS8 1UB, UK}
\selectlanguage{english}%

\begin{abstract}

Photonic qubits constitute a leading platform to disruptive quantum technologies due to their unique low-noise properties.  The cost of the photonic approach is the non-deterministic nature of many of the processes, including single-photon generation, which arises from parametric sources and negligible interaction between photons.  Active temporal multiplexing---repeating a generation process in time and rerouting to single modes using an optical switching network---is a promising approach to overcome this challenge and will likely be essential for large-scale applications with greatly reduced resource complexity and system sizes.  Requirements include the precise synchronization of a system of low-loss switches, delay lines, fast photon detectors, and feed-forward. Here we demonstrate temporal multiplexing of 8 `bins' from a double-passed heralded photon source and observe an increase in the heralding and heralded photon rates. This system points the way to harnessing temporal multiplexing in quantum technologies, from single-photon sources to large-scale computation.

\end{abstract}
\maketitle

\begin{figure}[t]
\begin{centering}
\includegraphics[scale=0.45]{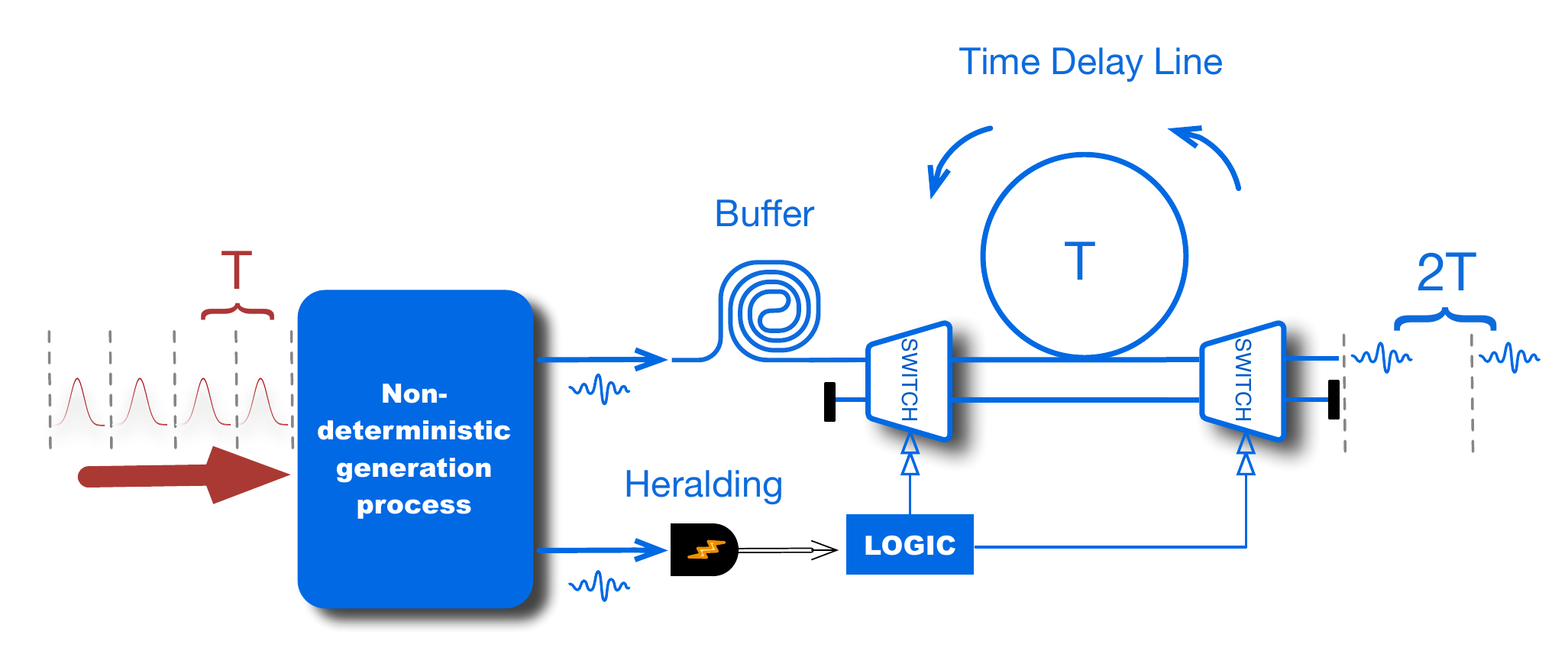}
\protect\caption{\textbf{Illustration of the principle of temporal multiplexing. } 
A non-deterministic generation process is repeated in time with period $T$; on heralded success, an active optical switching network and delay lines offset photons into output time bins spaced by an integer multiple of the input period and in sync with the system clock cycle.   %With enough input time bins and  a sufficiently low loss switching network, the generation probability per clock cycle is increased.
With a sufficiently low-loss switching network, the generation probability per clock cycle is increased.}\label{Fig:TimeMux}
\par\end{centering}
\end{figure}

\section{Introduction}

Preparation and manipulation of exotic quantum states of light are at the heart of quantum information science and technology \cite{pa-rmp-84-777}. 
A central challenge for photonic systems is the non-deterministic nature of the generation of such states, which arises when considering parametric photon sources \cite{Eisaman:2011cc}, such as those based on spontaneous parametric down-conversion (SPDC) or four-wave mixing (FWM), and from the negligible interaction between photons \cite{kn-nat-409-46}.  Parametric sources of single photons have been the workhorse for proofs of principle to date because they generate photons in pairs, enabling heralding in highly pure spatio-temporal-spectral modes \cite{Eisaman:2011cc}. However, parametric sources have a theoretical maximum heralding efficiency of 25\% \cite{ch-pra-85-023829}, sufficient for some communication and sensing applications, but short of the best known threshold for computation \cite{Br-prl-95-010501}.
As with other 
non-deterministic generation processes with heralded success signals, including fusion gates, for large-scale cluster states \cite{Br-prl-95-010501,gi-arx-1410-3720} and all-photonic quantum communication \cite{Azuma:2015}, and ballistic entangled state generation \cite{eventready,Varnava:2008bd,Cable:2007jr}, the success probabilities must be increased above relevant practical thresholds. 

A promising approach is to actively multiplex (MUX) these processes by operating several copies in parallel, such that the probability of at least one succeeding is high, followed by a low-loss switching network to route a successful output into the downstream system \cite{mi-pra-66-053805,wo-ol-32-18,Jennewein,gi-arx-1410-3720}. Spatial multiplexing of heralded photon sources \cite{mi-pra-66-053805,wo-ol-32-18,Jennewein,gi-arx-1410-3720,bo-arx-1409-5341,ch-pra-85-023829}, for example, has been successfully implemented with up to four heralded photon sources \cite{Ma:2011in,Collins:2013eu,Meany:2014jn,Xiong:2013}. Temporal multiplexing \cite{mo-pra-84-052326,sch-apb-11-447,PittmanSPIE2004} (see Fig. \ref{Fig:TimeMux}) would enable repeated use of the same physical process, reducing resources, system size and indistinguishability requirements, at the cost of introducing delay lines and reducing the system clock rate. Temporal  multiplexing has been proposed for single-photon \cite{mo-pra-84-052326, sch-apb-11-447,Rohde:2015,Francis-Jones:2015} and  entangled state generation  \cite{mo-pra-84-052326, gi-arx-1410-3720, PittmanSPIE2004,Kwiat2009}, as well as for photon memories \cite{PittmanSPIE2004}, boson sampling schemes \cite{Rohde:2014}, and universal quantum computation \cite{Rhode2015}.

Here, we demonstrate active temporal multiplexing and use it to improve the success probability of a heralded single-photon source. By combining temporal with spatial multiplexing using a double-passed heralded source, only a single physical source was used to enable hybrid spatial-temporal multiplexing of eight effective source repetitions. We show active temporal multiplexing of a source of periodic photons (previous demonstrations have shown passive temporal multiplexing \cite{Broome:2011el} and active storage, but not multiplexing, of photons from non-periodic sources using cavities \cite{pi-pra-66-042303,cavity}) to increase the heralded photon rate, for a fixed noise level, by up to 76\%  compared to the same source without multiplexing.

\begin{figure*}
\begin{centering}
\includegraphics[scale=.72]{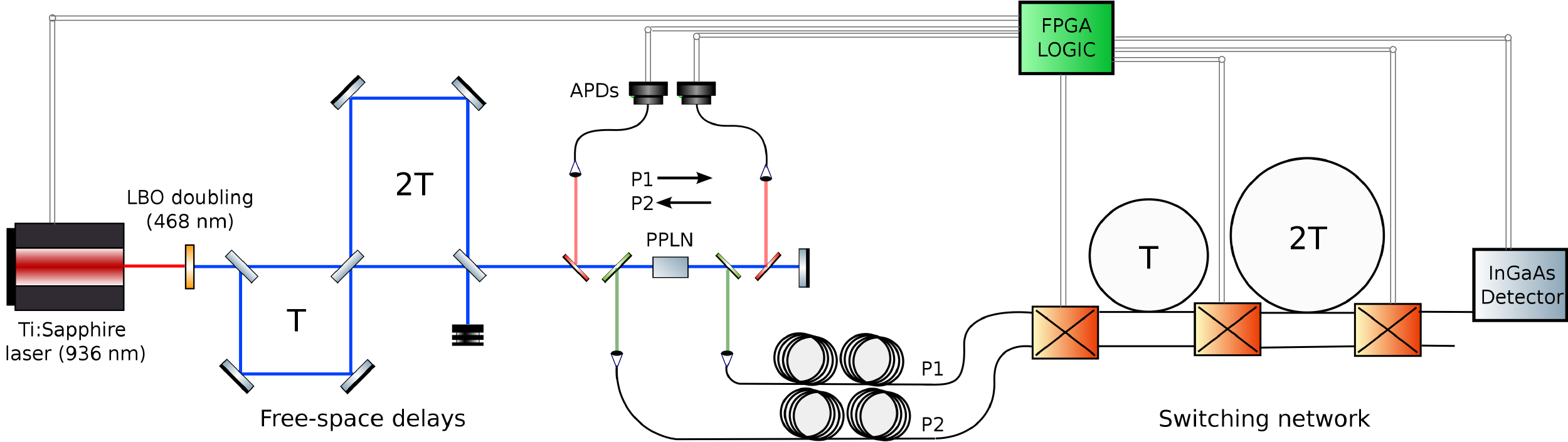}
\par\end{centering}
\centering{}\protect\caption{\textbf{Experimental set-up.}
Pulses from a femtosecond laser are upconverted using an LBO crystal,
split into four copies using free-space delay lines, and passed twice
through a PPLN crystal for down-conversion.  Following separation of the photon pairs and pump using filtering,  the heralding signals are analyzed by an oversampling
FPGA while the signal photons are stored in long fiber delays. The
FPGA configures the switching network to deliver the generated signal photons
into a single spatial and temporal mode. P1 and P2 indicate Pass 1 and Pass 2. }\label{exp}
\end{figure*}

\section*{Results}\label{Results}
\noindent \textbf{Principle of operation}

In a heralded single-photon source (HSPS), a pulsed laser pumps a nonlinear material, spontaneously %\ref{Fig:ExtrTheoEmission}
generating photon pairs, called signal and idler photons, in a fixed
time bin. Assuming spectral disentanglement \cite{Grice:2001jc}, the state after passing through the nonlinear material is given by an infinite superposition of Fock state pairs
\cite{Knight2004}:

\begin{equation}
\left|\psi\right\rangle =\sqrt{1-\left|\xi\right|^{2}}\left(\left|0\right\rangle _{i}\left|0\right\rangle _{s}+\sum_{n=1}^{\infty}\xi^{n}\left|n\right\rangle _{i}\left|n\right\rangle _{s}\right),\label{eq:SPDC state}
\end{equation}
where $i$ and $s$ are the idler and signal modes and $\xi$ is the
squeezing parameter determined by the pump power and the strength
of the nonlinearity. \foreignlanguage{english}{}Multi-photon pairs
generally result in detrimental effects in quantum circuits, necessitating
low squeezing parameters so that the single-pair term in (1) dominates. 

Filters are used to separate the signal and idler photon and the pump,
and a single-photon detector placed on the idler arm is used to herald
the presence of the signal photon. Under ideal conditions and with
number-resolving detectors, the theoretical maximum single-photon
emission probability of a HSPS is limited to 25\% \cite{ch-pra-85-023829}, due to
the presence of multi-photon pair terms in equation (1). While this
single-photon emission probability is sufficient for small-scale quantum
optics experiments, heralded sources by themselves are not sufficient
for scalable quantum technology \cite{bo-arx-1409-5341}. 

A temporal multiplexing technique (also referred to as time multiplexing) \cite{mo-pra-84-052326,sch-apb-11-447}, which uses a HSPS, optical switches, delay line loops, and electronics for feed-forward, can be used
to boost the single-photon emission probability while keeping the
multi-photon contamination low (Fig. \ref{Fig:TimeMux} shows the smallest example with one delay loop).  In this scheme, the HSPS is pumped $N$ times per clock cycle with
laser pulses spaced by time $T$.  Signal photons are stored in a long delay line buffer as detection signals from the idler arm are analyzed.  When a single photon is heralded in one of the $N$ time bins, a switching network composed of delay line loops (with lengths
of integer multiples of $T$) is driven into a configuration which offsets the photon into a single
spatial-temporal mode. If multiple photons are heralded in several input bins, the switching network automatically discards the extra photons by moving them into adjacent bins, thus ensuring that only a single photon is output in the time bin in sync with the system clock cycle.

With a sufficient number of time bins per clock cycle,
a single-photon pair will be produced in at least one of the time bins with high probability. The probability of heralded single-photon
emission from the multiplexed source is approximately (see Supplementary Material III for a detailed model): 

\selectlanguage{english}%
\begin{equation}
p_{single}^{MUX}=\left(1-\left(1-p_{trig}\right)^{N}\right)p_{single},
\end{equation}

where $p_{trig}$ is the probability that the HSPS triggers during one time bin
and $p_{single}$ is the probability that the triggered emission is a single photon after passing through the lossy switching network \cite{bo-arx-1409-5341}.  With
ideal operation and assuming a lossless switching network, 17 heralded
source repetitions enable a source with >99\% single-photon emission probability
\cite{ch-pra-85-023829}, and assuming realistically small losses, $\sim$8-16 heralded
source repetitions enable a near-deterministic source with low multi-photon contamination for large-scale applications
\cite{bo-arx-1409-5341}.  Even when considering heralded sources operating with efficiencies far below the theoretical maximum, as is the case with all parametric sources demonstrated to date, multiplexing can still be used to achieve an enhanced heralded single-photon emission probability per clock cycle for a fixed multi-photon contamination probability, offering the possibility of new classes of experiments in the near-term.

\begin{figure}
\begin{centering}
~\includegraphics[width=\columnwidth]{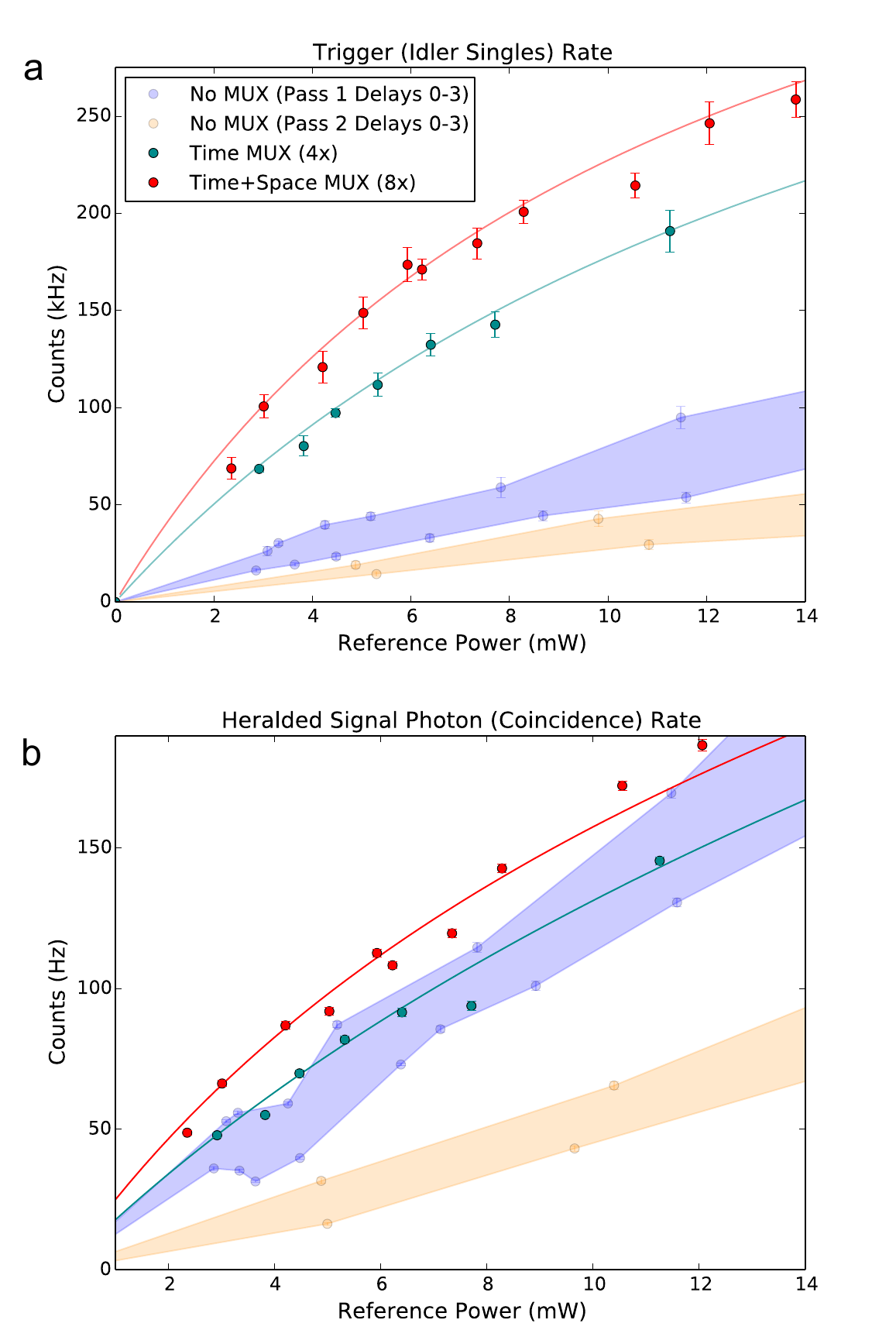}
\par\end{centering}

%\begin{centering}
%\includegraphics[width=\columnwidth]{coincidencesv5}
%\par\end{centering}
\centering{}\protect\caption{\textbf{Photon statistics from multiplexed and non-multiplexed sources.
a,} Triggering (idler singles) and \textbf{b,} and heralded signal photon (coincidence) rates vs.
reference laser pump power for the 8$\times$ multiplexed, 4$\times$ multiplexed,
and non-multiplexed sources. For clarity of presentation, data points from the non-multiplexed
sources are shown as linearly interpolated region plots encompassing
the range of data: blue (Pass 1, Delays 0-3) and orange (Pass 2, Delays
0-3). "Delays 0-3" refers to the four effective non-multiplexed sources passively delayed in time. Theory lines for the multiplexed sources are calculated from
measured heralded source parameters, measured switch loss, and extrinsic
loss effects. }\label{photon_statistic}
\end{figure}

\begin{figure}
\begin{centering}
\includegraphics[width=\columnwidth]{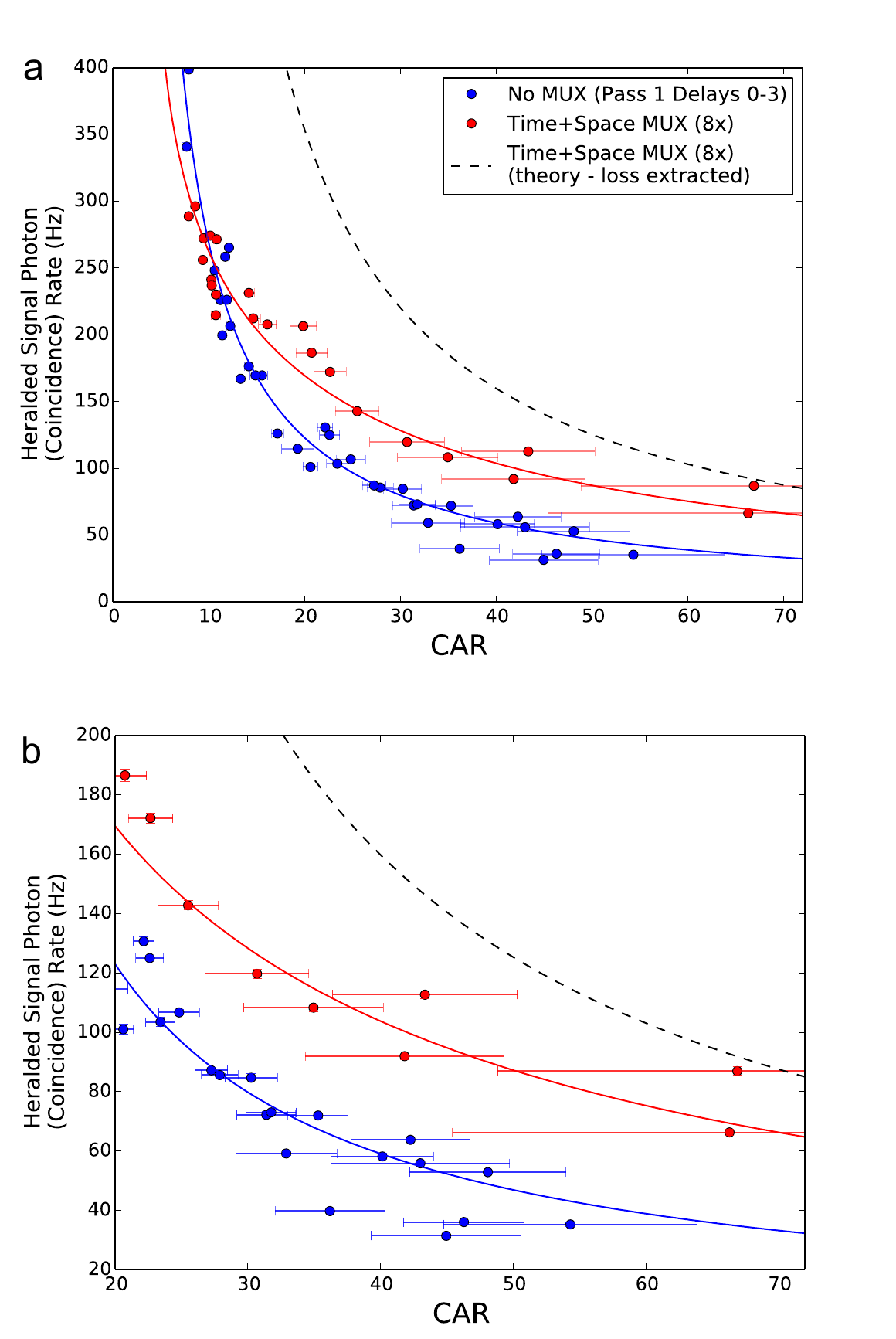}
\par\end{centering}

%\begin{centering}
%\includegraphics[width=\columnwidth]{CAR_graph_detailv6}
%\par\end{centering}

\centering{}\protect\caption{\textbf{Heralded signal photon (coincidence) rate vs. CAR for multiplexed
and non-multiplexed sources. a,} Shows the full data set, and \textbf{b,}
shows detail at low powers, where saturation effects due to electronics
are small. "Delays 0-3" refers to the four effective non-multiplexed sources passively delayed in time. Red points are for the 8$\times$ multiplexed source, and blue
points are for the non-multiplexed sources (Pass 1).  Solid lines are
based on a theory fit using measured parameters. Dashed line shows
a correction for extrinsic sources of loss based on the theory model. }\label{Fig:heraldedrate}
\end{figure}

In this work we implemented a hybrid scheme using two spatially
multiplexed sources fed into a temporal multiplexing set-up, doubling the number of effective source repetitions (Fig. \ref{exp}).
By using a return pass in the opposite direction through the same
nonlinear crystal, hybrid temporal and spatial multiplexing can be implemented
with only a single parametric source. This scheme enables an additional
enhancement to the single-photon emission probability without an additional
loss penalty on the generated photons, since the same depth of switches as the temporal multiplexing
scheme is used.

\noindent \textbf{Implementation}

Our experimental set-up (Fig. \ref{exp})
 uses a bulk periodically-poled lithium niobate (PPLN) down-conversion crystal phase-matched to produce idler photons at 671 nm and signal photons at 1547 nm from a pump laser at 468 nm. These wavelengths enable high-efficiency detection of the idler photon using silicon avalanche photodiodes (APDs) and low-loss transmission of the signal photon through switches and fiber delay lines. 

Each pulse from a pump laser with a 80 MHz repetition rate (12.5 ns pulse spacing)  is frequency doubled and split into four pulses spaced by  $\sim$3 ns using
a series of free-space delay lines constructed from beam-splitters and mirrors. The four pulses then pass through the PPLN crystal
and undergo collinear down-conversion, probabilistically creating
photon pairs in four time bins in the first spatial mode (referred to as Pass 1); a return pass
of the pump through the crystal, obtained by recycling the residual pump reflected from a mirror, creates four additional time bins
in a separate spatial mode (Pass 2).  For each pass, the four effective sources passively delayed in time are referred to as "Delays", e.g. Pass 1, Delay 3.  The spectra of the signal photons from different delays of the same pass were shown to have a high degree of similarity (mean of 97.9$\pm$1.8\%, see Supplementary material II), and moderate similarity between the two different passes (mean of 92.8$\pm$5.2\%, see Supplementary material II). Furthermore, an in-line polarizer was used to verify that the photons emitted from each source had identical polarization. 

The polarization-maintaining, active optical switching network (see Appendix A: Methods) is composed of low-loss fiber
switches ($\sim$1 dB loss per switch, 500 kHz maximum operation frequency) and two fiber delay loops matched to the free-space delay lines (see Fig. \ref{exp}) \cite{mo-pra-84-052326}.  Detection signal rising edges, which can fall in any of the four closely spaced time bins, are correctly discriminated using a fast oversampling Field Programmable
Gate Array (FPGA) (see Appendix A: Methods), which then configures the switches for feed-forward multiplexing
of the eight time and spatial modes. To avoid driving the switches
faster than their maximum operation frequency, an asynchronous "idle time"
of 2 $\mu$s was programmed into the FPGA to limit the rate of detected
heralding signals. During heralding detection, feed-forward processing,
and switch configuration, the signal photons are stored in long delay
lines of telecom fiber. Signal photons are detected using an InGaAs
detector, gated from the idler photon detection events.

\noindent \textbf{Measured photon statistics from multiplexed and non-multiplexed sources}

Photon counting statistics were collected for the eight non-multiplexed
sources and the multiplexed source. 
Triggering (idler singles) rates and heralded signal photon (coincidence) rates (see Appendix A: Methods)  are shown in Fig.~\ref{photon_statistic} (accidental rates are shown in Supplementary material I).  For all coincidence and accidental measurements, time-averaged dark counts, which had negligible effect at the measured rates, were subtracted from the totals. 
The data was taken for fixed "reference powers", defined as the average power of the pump laser in front of the PPLN crystal used to generate the photons in the {\em multiplexed} source. 
Then, by blocking all beam paths in the free-space delay lines except for one at a time, a fraction of the reference power was used to pump each of the
non-multiplexed sources individually ($\sim$25\% for Pass 1 delays and, due to power loss, $\sim$12.5\%
for Pass 2 delays). This procedure allows for a fair comparison of the coincidence, accidental, and triggering rates between the multiplexed source and its constituent non-multiplexed sources.

The triggering
rates were affected by saturation
loss caused by the large asynchronous "idle time" programmed into the FPGA to avoid triggering detection faster than the limited switch operation frequency (500 kHz). With the overall clock rate of the source set by the 80 MHz repetition rate of the pump laser, the multiplexed source suffered from greater saturation effects at high powers due to this deadtime than the less-deterministic non-multiplexed sources, as can be seen in the sharp bending
of the data away from a linear trend as the reference power is increased in Fig. \ref{photon_statistic}a.

The data was found to be in excellent agreement with our model of the
non-multiplexed and multiplexed sources (see Supplementary material III), as shown in Fig. \ref{photon_statistic}.
According to the fit, the 4$\times$ time multiplexed source (composed of all Pass 1 delays)
showed an increase of up to 175\% in the triggering rate, for a fixed reference power, compared to the
most efficient non-multiplexed source (Pass 1, Delay 3) and the 8$\times$ time and space multiplexed source (composed of both Passes, all delays) showed
up to a 290\% increase (see Fig. \ref{photon_statistic}a).  At low reference powers (2.5-10.5 mW), the difference in coincidence rates for the 4$\times$ multiplexed source was not significant compared to the non-multiplexed
source with the highest coincidence rates, while the 8$\times$ multiplexed source showed a higher coincidence rate than any individual non-multiplexed source, and a 17-69\% higher rate compared to the mean from the non-multiplexed
sources (Pass 1 only)  (see Fig. \ref{photon_statistic}b). At high powers, the rates of coincidences for the multiplexed sources were suppressed due to the saturation of triggering events, as predicted by our model (Supplementary material III).

%\noindent \textbf{Coincidence-to-accidental ratio (CAR)}

Although the triggering and coincidence rate statistics provide evidence that our set-up implements temporal multiplexing, the key measure of performance for our multiplexed source is the heralded photon rate for a fixed coincidence-to-accidental ratio (CAR).  The CAR serves as a measure of noise due to single-photon emission and multi-photon contamination probabilities; this measure of noise cannot be inferred from coincidence rates alone. The heralded photon rates for fixed CAR for the 8$\times$ multiplexed and non-multiplexed sources are plotted in Fig. \ref{Fig:heraldedrate} (data from the 4$\times$ multiplexed
source is shown in  Supplementary material I). 
 For a fixed CAR, in the regime where saturation
effects are small, the 4$\times$ multiplexed
source did not show a significant increase in the heralded photon
rate, 
and was limited mainly
by the loss of the switching network ($\sim$4 dB loss for each path).  However, the 8$\times$ multiplexed source exhibited an increased
heralded photon rate between 33-59\%, for the same CAR, over the best non-multiplexed
source, and between 47-76\% over the mean from the non-multiplexed
sources (Pass 1 only), demonstrating a direct improvement (Fig. \ref{Fig:heraldedrate}b). 
In our model we corrected for the effects of extrinsic loss on the rate of heralded photon production (dashed line Fig. \ref{Fig:heraldedrate}a, see Methods and Supplementary material III), indicating a potential improvement of $\sim$114\% for a wide range of CAR values compared to the expected heralded photon rate from the non-multiplexed sources (also with extrinsic loss removed).  This enhancement can be mapped to a comparable increase in the single-photon emission probability for a fixed multi-photon emission probability (see Supplementary material IV).

\section*{Discussion}\label{discussion}

We demonstrated temporal and spatial multiplexing of eight photon bins in a hybrid setup to enhance the heralded photon emission statistics compared to non-multiplexed sources.  Although our demonstrated improvement was limited by the maximum operation frequency of the switches, we note that even low rate, high efficiency multiplexed sources will likely find applications in the near term, due to the increased single-photon generation probability per clock cycle.  Therefore, a possible solution is to use a pulse picker to limit the repetition rate of the laser source to the maximum repetition rate of the switches. Ultimately, multiplexed sources with the highest single-photon emission rates will require the development of a lower-loss, high-speed optical switch (recent, promising prototypes include Kerr effect \cite{Rambo:2013} and electro-optic based \cite{Rao:2015} switches). 

Demonstrating Hong-Ou-Mandel interference between two independent multiplexed sources, which has not yet been demonstrated in any active multiplexing implementation, is the next step in order to verify that photon purity is preserved. In our multiplexed source, photons from different delays had similar spectral properties, identical polarizations, and similar source efficiencies and couplings. Our setup of adjustable free-space delays allows for the fine tuning of the temporal delays of the photons to within a coherence length, indicating that photon interference between two of our multiplexed sources, up to the intrinsic limit of the PPLN sources themselves, should be possible.   Improved multiplexing components and sources will further enable the temporal multiplexing of more complex generation processes, such as fusion gates \cite{Br-prl-95-010501,gi-arx-1410-3720} or ballistic entangled state generation \cite{eventready,Varnava:2008bd,Cable:2007jr}, where phase stability will be essential. 

Temporal multiplexing techniques will almost certainly
be required in future large-scale quantum photonic circuits in order to substantially
reduce resource requirements. Furthermore, hybrid temporal and spatial
multiplexing techniques will be important in order to optimize tradeoffs
between spatial footprint and system clock rate.  Integrated photonic components, including sources (e.g. \cite{Silverstone:2014fu}),  switches (e.g. \cite{Lacava:2013ey}), filters (e.g. \cite{Jeong:2013es}),
delay lines (e.g. \cite{Lee:2012is}), and detectors (e.g. \cite{Marsili:2013fs}), are under development. 
Scaling down our set-up
to a fully integrated photonic chip with low-loss components will
enable a temporal multiplexing template capable of realizing
new classes of quantum information experiments and technology.

\section*{APPENDIX A: METHODS}\label{METHODS}

\noindent \textbf{Experimental set-up} 

A mode-locked, Ti:Sapphire laser ("Tsunami", Spectra Physics) produced
$\sim$150 femtosecond pulses at 936 nm; a LBO crystal (Newlight)
was used to frequency convert to 468 nm.  To enable low-loss,
near 50-50 splitting of the pulsed pump beam, laser line non-polarizing
beamsplitters (Newport) were used in the free space delays. The PPLN
crystal (Covesion) was 3 mm long and phase-matched at 110\textdegree{}
C using an oven and temperature controller. Dichroic mirrors (Semrock)
were used to separate the signal and idler photons from the pump,
and Pellin-Broca prisms were used for further spatial filtering. A
bandpass filter centered at 671 nm (Semrock) was used on the idler arms
of each pass for further filtering.  

Polarization-maintaining switches (Agiltron, $\sim$1
dB loss per switch, 500 kHz max operation frequency) were based on an electro-optic
material. Standard telecom fiber was used for the long delay buffer ($\sim$200 m, $\sim$90\% transmission) and polarization-maintaining, low-dispersion fiber (Corning) was used for the variable delay line loops ($\sim$0.65 m and 1.30 m, $\sim$95\% transmission). Fiber polarization controllers (Fiberpro) were used before the polarization-maintaining switching network to match the polarizations of photons from the two passes.  To enable reliable comparison between multiplexed and non-multiplexed
source measurements, a MEMS switch with nearly balanced loss was used
to route the photons into or around the multiplexing switch network. 
  
\noindent\textbf{Photon detection} 

Idler (triggering) photons were detected using silicon avalanche photodiodes (APDs) (PerkinElmer).  Pump leakage and dark counts were found to be negligible on the idler arms.  

Idler photon detection signals were discriminated with an oversampling FPGA (Xilinx Spartan 6) using internal delays and a 80 MHz, "locked-to-clock" reference input from the Ti:Sapphire laser.  The FPGA was designed so that after signal detection, an "idle time" of 2 $\mu$s became active to avoid further detection. For every detected signal (regardless of time bin), a gating signal was output with a constant delay with reference to the original input clock, to ensure correct heralding of the temporally multiplexed photons.  The total (unoptimized) internal delay of the  FPGA logic was $\sim$60 ns.

Signal photons were detected using InGaAs detectors (ID Quantique).  Coincidence counts, joint detection between idler and signal photons
from paired generation events, were collected using gated detection
of the signal photon from idler detection signals from the FPGA. Accidental counts,
joint detection between idler and gated signal photons from unpaired
generation events, were then collected by shifting the temporal delay
of the FPGA input clock by a multiple of the clock cycle. Pump leakage
in the signal arms was found to be negligible at the measured powers.
Dark counts detected by the InGaAs detectors in gated mode were measured
by blocking the signal arm path; these time-averaged counts were then subtracted
from the count totals. The gate width used was 1.8 ns.

\noindent\textbf{Extrinsic sources of loss}

Extrinsic source of loss in the setup include: 1) Loss due to measurement
apparatus. A small amount of extra loss (4\%) on the multiplexed source
was due to asymmetric loss of the MEMS switch used to switch between
multiplexing and non-multiplexing channels for measurement. 2) Loss
due to the deadtime of available electronic amplifiers. The two electronic
amplifiers used to amplify the signal from the APD and into the FPGA
have deadtimes of \ensuremath{\sim} 0.1 $\mu$s, resulting in missed
pulses from the APD. In principle, much faster amplifiers with negligible
deadtimes can be used to eliminate this source of loss. 3) Loss due
to the limited switch repetition rate (500 kHz). An asynchronous "idle time"
of 2 $\mu$s was programmed into the FPGA to avoid driving the switches
faster than their 500 kHz maximum operation frequency. The switch repetition rate
is set by the switch driver board; the switches themselves have
a faster intrinsic rise and fall time of 300 ns (\ensuremath{\sim}
3 MHz).

\section*{Acknowledgements}

We thank Xiao Ai, Daryl Beggs, Allison Rubenok, Gary Sinclair, Ivo Straka, Jianwei Wang, Andrew Young, and Xiou-Qi Zhou for useful discussions and assistance. This work was supported by EPSRC, ERC, PICQUE, BBOI, US Army Research Office (ARO) Grant No. W911NF-14-1-0133, U.S. Air Force Office of Scientific Research (AFOSR). J.L.O\textquoteright B. acknowledges a Royal Society Wolfson Merit Award and a Royal Academy of Engineering Chair in Emerging Technologies. G.D.M. acknowledges the FP7 Marie Curie International Incoming Fellowship scheme.

%\section*{Contributions}

%G.J.M. and D.B. designed the experiment. G.J.M., R.S., J.M., E.H., E.M. carried out the experiment. G.D.M., M.G.T., and J.L.O. supervised the work. G.J.M. and R.S. wrote the manuscript with input from all authors. 

\bibliographystyle{unsrt}

%\bibliography{references} 

\clearpage

%%%%%%%%%% Merge with supplemental materials %%%%%%%%%%

\pagebreak
\newpage
\widetext
\begin{center}

\section*{\large Supplementary Material: Active Temporal Multiplexing of Photons}
\end{center}
%%%%%%%%%% Merge with supplemental materials %%%%%%%%%%
%%%%%%%%%% Prefix a "S" to all equations, figures, tables and reset the counter %%%%%%%%%%
\setcounter{equation}{0}
\setcounter{figure}{0}
\setcounter{table}{0}
\setcounter{page}{1}
\makeatletter
\renewcommand{\theequation}{S\arabic{equation}}
\renewcommand{\thefigure}{S\arabic{figure}}
\renewcommand{\bibnumfmt}[1]{[S#1]}
\renewcommand{\citenumfont}[1]{S#1}
\onecolumngrid

\section{Accidentals and CAR Plots}
\label{sec:Accidental-and-CAR-Plots}

A plot of accidental rates against reference power for the multiplexed
and non-multiplexed sources is shown in Fig. S1a. A plot of coincidence
rates against CAR, including data from the 4$\times$ multiplexed
source and from Pass 2 is shown in Fig. S1b.

\section{Source Details and Spectral Data}
\label{sec:source}

\begin{figure}[b!]

\begin{centering}
\includegraphics[scale=0.82]{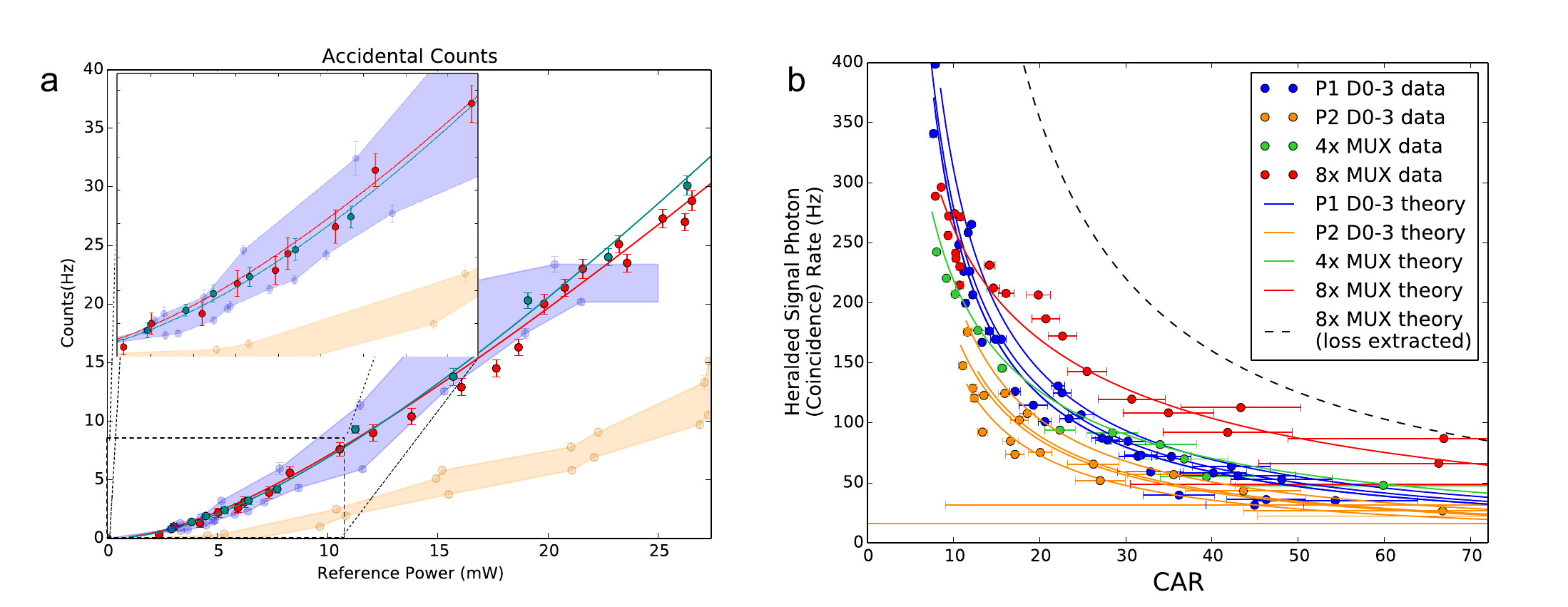} 
\par\end{centering}

\centering{}\protect\protect\caption{\textbf{Accidental and CAR plots. a, }Accidental rates vs. reference
laser pump power for the 8x multiplexed (red), 4x multiplexed (green),
and non-multiplexed sources. For clarity, data points from the non-multiplexed
sources are shown as linearly interpolated region plots encompassing
the range of data: blue (Pass 1, Delays 0-3) and orange (Pass 2, Delays
0-3). P1 D0 indicates Pass 1, Delay 0, and similarly for the other
labels. Theory lines for the multiplexed sources are calculated from
measured heralded source parameters, measured switch loss, and extrinsic
loss effects. Inset shows detail at low powers, where saturation effects
due to electronics are small.\textbf{ b, Heralded signal photon (coincidence)
rate vs. CAR for multiplexed and non-multiplexed sources. }Red points
are for the 8x multiplexed source, green points are for the 4x multiplexed
source, blue points are for the non-multiplexed sources (Pass 1),
and orange points are for the non-multiplexed sources (Pass 2). P1
D0 indicates Pass 1, Delay 0, and similarly for the other labels.
Solid lines are based on a theory fit using measured parameters. Dashed
line shows a correction for extrinsic sources of loss based on the
theory model.}

\label{Fig:accrates} 
\end{figure}

\begin{figure*}
\begin{centering}
\includegraphics[scale=0.45]{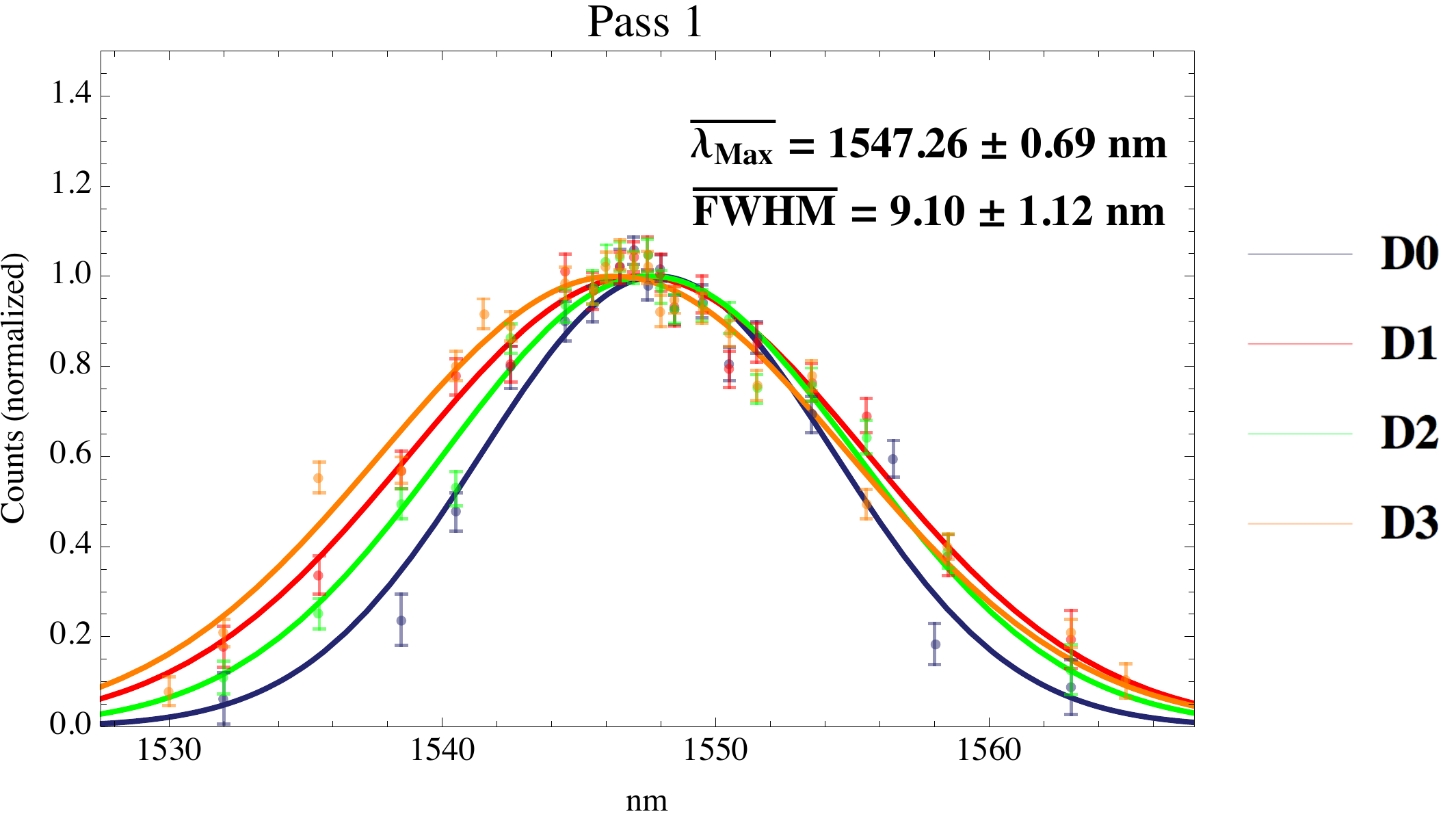} 
\par\end{centering}

\begin{centering}
\includegraphics[scale=0.45]{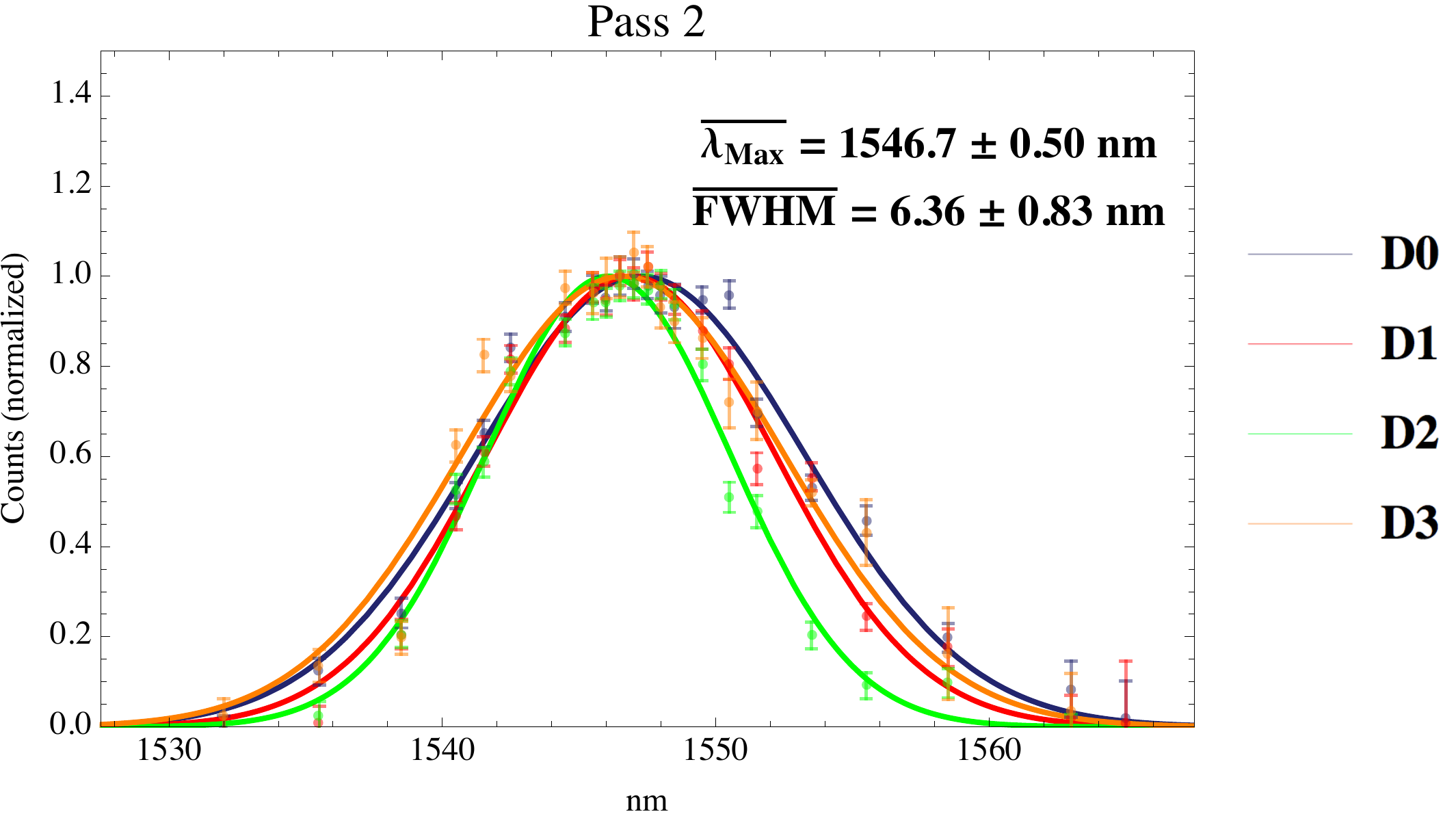} 
\par\end{centering}

\centering{}\protect\protect\caption{\textbf{Source spectral characterization.} Normalized spectral data
from signal photons from both passes. D0 indicates Delay 0, and similarly
for the other labels. $\bar{\lambda}_{max}$ is mean peak position
and $\overline{\textrm{FWHM}}$ is mean full-width half maximum for
all delays in a pass. Curves are best fits to a Gaussian.}

\label{Fig:Spectra} 
\end{figure*}

\begin{table}

\begin{centering}
\begin{tabular}{|c|c|c|c|c|}
\hline 
\textbf{Pass 1}  & Delay 0 & Delay 1 & Delay 2 & Delay 3\tabularnewline
\hline 
Delay 0 & 1.0 & 0.966 & 0.988 & 0.952\tabularnewline
\hline 
Delay 1 &  & 1.0  & 0.994 & 0.997\tabularnewline
\hline 
Delay 2 &  &  & 1.0 & 0.986\tabularnewline
\hline 
Delay 3 &  &  &  & 1.0\tabularnewline
\hline 
\end{tabular}\protect
\par\end{centering}

~

~

\begin{centering}
\begin{tabular}{|c|c|c|c|c|}
\hline 
\textbf{Pass 2 } & Delay 0 & Delay 1 & Delay 2 & Delay 3\tabularnewline
\hline 
Delay 0 & 1.0 & 0.991 & 0.946 & 0.997\tabularnewline
\hline 
Delay 1 &  & 1.0  & 0.979 & 0.993\tabularnewline
\hline 
Delay 2 &  &  & 1.0 & 0.960\tabularnewline
\hline 
Delay 3 &  &  &  & 1.0\tabularnewline
\hline 
\end{tabular}\protect
\par\end{centering}

~

~

\centering{}%
\begin{tabular}{|c|c|c|c|c|}
\hline 
 & P1 D0 & P1 D1 & P1 D2 & P1 D3\tabularnewline
\hline 
P2 D0 & 0.997 & 0.979 & 0.925 & 0.993\tabularnewline
\hline 
P2 D1 & 0.947 & 0.903 & 0.828 & 0.941\tabularnewline
\hline 
P2 D2 & 0.967 & 0.936 & 0.877 & 0.969\tabularnewline
\hline 
P2 D3 & 0.938 & 0.893 & 0.819 & 0.934\tabularnewline
\hline 
\end{tabular}\protect\protect\caption{\textbf{\label{tab:Table-showing-indistinguishabili}Tables showing
upper bounds on indistinguishability parameters between photons from
different passes and delays.} Top table shows upper bounds on indistinguishability
parameters from photons from different delays in Pass 1. Middle table
shows upper bounds on indistinguishability parameters from photons
from different delays in Pass 2. Bottom table shows upper bounds on
indistinguishability parameters from photons from delays from different
passes. "P" and "D" refer to the pass and delay number, respectively.}

\end{table}

Although Pass 2 should ideally be identical in operation to Pass 1,
it differs from Pass 1 in two aspects. First, the laser pump power
passing through the down-conversion crystal has been reduced by $\sim$50\%
compared to the first pass, due to loss from the crystal without anti-reflection
coating ($\sim$15\% pump loss per facet) and additional loss from
dichroic mirrors used for filtering. Second, Pass 2 is affected by
unwanted ``back-reflected\textquotedblright{} photons from the Pass
1, again due to a lack of anti-reflection coating and the cut of the
crystal. These back-reflected photons result in false heralding events,
degrading the quality of the multiplexed source. By optimizing the
position of the crystal, the back-reflected photon contamination in
Pass 2 can be minimized ($\sim$20\% of the total counts on the idler
arm and $\sim$8\% of the total counts on signal arm). This
results in a slight change in the CAR, although the larger effect
is additional saturation loss (described in the main text) due to
the increased triggering rate. These features were taken into account
during the modelling and analysis of results (Supplementary material
\ref{sec:Theoretical-Model}).

Spectral data from all eight effective sources was collected
by detecting signal photons using InGaAs detectors after passing through
a tunable filter (Yenista), with time-averaged dark counts subtracted
(Fig. S2). To quantify the similarity between photons from different
passes and delays, upper bounds on the indistinguishability parameters
$\gamma$ \cite{Rhode06} between each pair of sources were calculated using
integral overlaps of the fit spectral amplitude functions (Table \ref{tab:Table-showing-indistinguishabili}).
The integral overlap formula is

\begin{equation}
\gamma=\left|\int_{-\infty}^{\infty}\psi_{A}\left(\omega\right)\psi_{B}^{*}\left(\omega\right)d\omega\right|^{2},
\end{equation}

where $\psi_{A}\left(\omega\right)$ and $\psi_{B}\left(\omega\right)$
are the spectral wavefunctions of the photons from two different effective
sources. $\gamma$ obeys $0\leq\gamma\leq1$, where $\gamma=1$ corresponds
to perfect indistinguishability and $\gamma=0$ corresponds to perfect
distinguishability. We emphasize that this is only an upper bound
on indistinguishability, and does not consider other degrees of freedom
or joint spectral entanglement between signal and idler photons. Between
different delays from Pass 1, the mean overlap was 0.980$\pm$.017,
and between different delays from Pass 2, the mean overlap was 0.978$\pm$.020.
This similarity can likely be further improved by passing the pump
beam through a single mode filter such as a pin hole setup before
the PPLN crystal, eliminating possible angular and intensity deviations
in the pump beams for different delays. Between delays from different
passes, the mean overlap was 0.928$\pm$.052. The difference in spectra
can likely be explained by slight deviations in the Pellin-Broca filtering
and coupling setups for the different passes, a feature that can affect
any two independent sources and is not inherent to the multiplexing
setup. Careful alignment optimizations, or further narrowband filtering,
can help eliminate this source of distinguishability. 

\section{Theoretical Model}
\label{sec:Theoretical-Model}

\subsection{Heralded single-photon source (HSPS) Model}

We assume that strong spectral filtering on the idler arm and moderate
filtering on the signal arm are sufficient such that the states produced
by our sources are close to an idealized two-mode squeezed state with
disentangled joint spectra. Each source is characterized by three
parameters: 1) an input power seed $P_{seed}$ which relates the amount
of input pump power to a reference photon generation probability from
the crystal ($p_{pair}=0.1)$. 2) Idler transmission $\eta_{i}$,
which includes all sources of loss on the idler photon, such as from
filters, coupling, and detector inefficiencies. The effect of nonlinear
loss due to saturation effects from electronics and deadtime are treated
in the next subsections. 3) Signal transmission $\eta_{s}$ which
includes all sources of loss on the signal photon.

The relation between pump input power and effective squeezing parameter
$\xi$ for a down-conversion source is given by

\begin{equation}
\xi=\tanh\left(c\sqrt{P}\right),
\end{equation}

where $P$ is the input power in units of mW and $c$ is a coupling
constant in units of mW$^{-1/2}$. The squeezing parameter $\xi_{seed}$
corresponding to $p_{pair}=0.1$ is $\xi_{seed}\approx.335715$, allowing
for the extraction of the coupling constant $c$ given a seed power
$P_{seed}$. With this coupling constant, the squeezing parameters
corresponding to any input power can be found.

Given a squeezing parameter $\xi$, idler transmission $\eta_{i}$
, and signal transmission $\eta_{s}$, the probability for the idler
(herald) arm to trigger using threshold detectors is given by \cite{bo-arx-1409-5341}:

\begin{equation}
p_{trig}^{i}=\frac{\left|\xi\right|^{2}\eta_{i}}{1-\left|\xi\right|^{2}\left(1-\eta_{i}\right)}.\label{eq:ptrig}
\end{equation}

Given that the herald has triggered, the probability for a single-photon
emission to lead to a detection event on the signal arm is

\begin{equation}
p_{single}^{s}=\left(1-\left|\xi\right|^{2}\right)\eta_{s}\frac{\left[1-\left(\left|\xi\right|^{2}\left(1-\eta_{s}\right)\right)^{2}\left(1-\eta_{i}\right)\right]\left[1-\left|\xi\right|^{2}\left(1-\eta_{i}\right)\right]}{\left[1-\left|\xi\right|^{2}\left(1-\eta_{s}\right)\right]^{2}\left[1-\left|\xi\right|^{2}\left(1-\eta_{s}\right)\left(1-\eta_{i}\right)\right]^{2}}.\label{gather:psingle}
\end{equation}

Given that the herald has triggered, the probability for a multi-photon
emission to lead to a detection event on the signal arm is

\begin{equation}
p_{multi}^{s}=\frac{Z_{TD}}{p_{trig}^{i}}-p_{single}^{s},\label{eq:p_multi_TD}
\end{equation}

with
\begin{equation}
Z_{TD}=\left(1-\left|\xi\right|^{2}\right)\left|\xi\right|^{2}\left(\frac{1}{1-\left|\xi\right|^{2}}+\frac{\left(1-\eta_{s}\right)\left(1-\eta_{i}\right)}{1-\left|\xi\right|^{2}\left(1-\eta_{s}\right)\left(1-\eta_{i}\right)}-\frac{\left(1-\eta_{i}\right)}{1-\left|\xi\right|^{2}\left(1-\eta_{i}\right)}-\frac{\left(1-\eta_{s}\right)}{1-\left|\xi\right|^{2}\left(1-\eta_{s}\right)}\right).
\end{equation}

Using these expressions we can find the probability of coincidence
and accidental detection, and CAR (coincidence-to-accidental ratio).

\subsection{Coincidences, accidentals, and CAR}

We can find the probability of coincidence detection using threshold
detectors, assuming the contribution from dark counts has been subtracted
and pump leakage is negligible.

The probability of a coincidence is given by

\begin{equation}
P_{C}=p_{trig}^{i}\left(p_{single}^{s}+p_{multi}^{s}\right).
\end{equation}

When seeded with a pump laser with repetition rate $R$, the expected
coincidence rate is

\begin{equation}
R_{C}=R\times P_{C}.
\end{equation}

An accidental occurs when a coincidence between an idler and signal
photon generated from two different pump pulses occurs, assuming the
subtraction of dark counts and negligible pump leakage.

Without triggering from a herald, the signal photon trigger probability
is given by

\begin{equation}
p_{trig}^{s}=\frac{\left|\xi\right|^{2}\eta_{s}}{1-\left|\xi\right|^{2}\left(1-\eta_{s}\right)}.
\end{equation}

Then a good approximation to the accidental probability is given by

\begin{equation}
P_{A}=p_{trig}^{i}p_{trig}^{s}.
\end{equation}

The rate of accidental detection events is then

\begin{equation}
R_{A}=R\times P_{A}.
\end{equation}

The expected CAR, coincidence to accidental ratio, is then

\begin{equation}
CAR=R_{C}/R_{A}.
\end{equation}

\subsection{Electronics saturation}

In practice, saturation effects due to electronics and detector deadtimes
will further affect the source performance. Assuming a detector has
a deadtime of $d$, and that events are approximately uniformly distributed,
the true rate of counts $T$ from a detected rate of counts $D$ can
be approximated as \cite{Knoll1989}

\begin{equation}
T=\frac{D}{1-Dd}.
\end{equation}

Conversely, given an expected rate of counts $T$, the detected rate
of counts will be

\begin{equation}
D=\frac{T}{dT+1}.
\end{equation}

By applying these equations several times, the effect of several electronic
deadtimes in series can be modelled. Most of the saturation effects
in the experiment affect the idler arm, so these equations can be
used to find an effective $p_{trig}^{i}$ probability. Then, an effective
$\eta_{i}$ which includes loss due to saturation can be determined
using equation \ref{eq:ptrig}.

\subsection{HSPS Model fitting to data: Pass 1}

Pass 1 has four time bins passing through the down-conversion crystal,
which we will call Delay 0, Delay 1, Delay 2, and Delay 3. We will
label these sources with the tuple $\left(pass,delay\right)$ corresponding
to the pass and delay associated with the state source, for example
Source $\left(1,3\right)$ identifies the source from Pass 1 Delay
3.

Although in the ideal case each source would be exactly the same,
in practice each source is slightly different: due to slight imperfections
in alignment and optics, each source may feature different relative
pump input powers, pump coupling, and different signal and idler loss
rates.

Using the above model of the HSPS, a numerical optimization was performed
to find the optimal input power seed $P_{seed}$, idler transmission
value $\eta_{i}$, and signal transmission value $\eta_{s}$ to match
recorded values of heralding triggers, coincidences, and accidentals.
We will label these parameters for each source with the corresponding
tuple: $\eta_{i}^{(pass,delay)}$, $\eta_{s}^{(pass,delay)}$, $P_{seed}^{(pass,delay)}$.
The models which maximized the mean $R^{2}$ statistic for trigger counts,
coincidences, and accidentals were found for all delays. These are
shown in Table \ref{Tab:Spectra}.

\begin{table}
\centering{}%
\begin{tabular}{|c|c|c|c|c|}
\hline 
Source  & $\eta_{i}$  & $\eta_{s}$  & $P_{seed}$ (mW)  & $\overline{R^{2}}$\tabularnewline
\hline 
Pass 1 Delay 0  & 0.015  & 0.0019  & 5.2  & 0.993\tabularnewline
\hline 
Pass 1 Delay 1  & 0.015  & 0.0019  & 6.8  & 0.990\tabularnewline
\hline 
Pass 1 Delay 2  & 0.016  & 0.0021  & 5.6  & 0.988\tabularnewline
\hline 
Pass 1 Delay 3  & 0.017  & 0.0018  & 4.6  & 0.993\tabularnewline
\hline 
\end{tabular}\protect\protect\caption{\textbf{Table showing source parameters for Pass 1.} Source parameters
were determined using a numerical optimization fit to the model. $\overline{R^{2}}$
is the mean $R^{2}$ from the triggering rate, coincidence, and accidental
fits.}
\label{Tab:Spectra} 
\end{table}

\subsection{Time Multiplexed Source Model: Pass 1}

The time multiplexed source is pumped by laser pulses which have been
split into four pulses using free-space delay lines (Main text: Fig.
2), so the time multiplexed source model is constructed from the four
HSPSs models from section D. Each HSPS in the time multiplexed source
is pumped with a fraction of the total power used to pump the multiplexed
source; the experimentally measured values were 0.2375, 0.2693, 0.2258,
0.2586 for Delay 0,1,2, and 3, respectively. The model for the time
multiplexed source follows that of \cite{bo-arx-1409-5341}, slightly
modified to account for each time bin source being slightly different.
In this section we neglect saturation effects; the same methods from
section C can then be applied to the expected detection events to
derive detection events with saturation.

The probability for the multiplexed source to trigger is given by

\begin{equation}
p_{trig}^{MUX,1}=1-\left(1-p_{trig}^{i,(1,0)}\right)\left(1-p_{trig}^{i,(1,1)}\right)\left(1-p_{trig}^{i,(1,2)}\right)\left(1-p_{trig}^{i,(1,3)}\right).
\end{equation}

When seeded with a pump laser with repetition rate $R$, the heralding
rate is then

\begin{equation}
R_{trig}^{MUX,1}=R\times p_{trig}^{MUX,1}.
\end{equation}

Let $\eta_{sw}^{(pass,delay)}$ correspond to the transmission due
to the switching network and variable delay lines for source $\left(pass,delay\right)$.
Then the probability the heralded state is a single photon emission
from source $\left(pass,delay\right)$, including the effect of the
switching network with variable delay lines is $p_{single}^{s',(pass,delay)}$,
given by replacing each instance of $\eta_{s}$ in equation \ref{gather:psingle}
with $\eta_{s}\eta_{sw}^{(pass,delay)}$. Similarly, the probability
the heralded state is multi-photon emission from source $\left(pass,delay\right)$
is $p_{multi}^{s',(pass,delay)}$, is given by replacing each instance
of $\eta_{s}$ in equation \ref{eq:p_multi_TD} with $\eta_{s}\eta_{sw}^{(pass,delay)}$.

The coincidence probability is

\begin{equation}
P_{C}^{MUX,1}=P_{C}^{\left(1,0\right)'}+\left(1-p_{trig}^{i,\left(1,0\right)}\right)\left(P_{C}^{\left(1,1\right)'}+\left(1-p_{trig}^{i,\left(1,1\right)}\right)\left(P_{C}^{\left(1,2\right)'}+\left(1-p_{trig}^{i,\left(1,2\right)}\right)P_{C}^{\left(1,3\right)'}\right)\right).
\end{equation}

where $P_{C}^{\left(1,delay\right)'}=p_{trig}^{i,\left(1,delay\right)}\left(p_{single}^{s^{'}\left(1,delay\right)}+p_{multi}^{s^{'}\left(1,delay\right)}\right)$.

When seeded with a pump laser with repetition rate $R$, the expected
coincidence rate is

\begin{equation}
R_{C}^{MUX,1}=R\times P_{C}^{MUX,1}.
\end{equation}

The accidental probability is
\begin{equation}
P_{A}^{MUX,1}=p_{trig}^{i,\left(1,0\right)}p_{trig}^{s',\left(1,0\right)}+\left(1-p_{trig}^{i,\left(1,0\right)}\right)\left(p_{trig}^{i,\left(1,1\right)}p_{trig}^{s',\left(1,1\right)}+\left(1-p_{trig}^{i,\left(1,1\right)}\right)\left(p_{trig}^{i,\left(1,2\right)}p_{trig}^{s',\left(1,2\right)}+\left(1-p_{trig}^{i,\left(1,2\right)}\right)p_{trig}^{i,\left(1,3\right)}p_{trig}^{s',\left(1,3\right)}\right)\right).
\end{equation}

The rate of accidental detection events is then

\begin{equation}
R_{A}^{MUX,1}=R\times P_{A}^{MUX,1}.
\end{equation}

The expected CAR, coincidence-to-accidental ratio, is then

\begin{equation}
CAR^{MUX,1}=R_{C}^{MUX}/R_{A}^{MUX}.
\end{equation}

Using measured values of switching loss and electronics saturation,
the model was found to be in good agreement with the data (Main text:
Fig. 3).

\subsection{HSPS Model fitting to data: Pass 2}

Pass 2 has two features different from Pass 1 (Supplementary material
\ref{sec:source}). We used
a simple model describing the effect of the back-reflected photon
contamination. We assume the back-reflected photons on the signal
arm are negligible. On the idler arm, we assume that every delay has
the same probability of triggering due to a back-reflected photon
as a fraction $f$ of the ``true'' triggering probability.

\begin{equation}
p_{trig\, back}^{i,\left(2,delay\right)}=p_{trig\, true}^{i,\left(2,delay\right)}\times f.
\end{equation}

Then the probability of ``correctly'' triggering is

\begin{equation}
p_{trig,correct}^{i,\left(2,delay\right)}=p_{trig\, true}^{i,\left(2,delay\right)}(1-p_{trig\, back}^{i,\left(2,delay\right)})+p_{trig\, true}^{i,\left(2,delay\right)}(p_{trig\, back}^{i,\left(2,delay\right)}),
\end{equation}

and the probability of ``incorrectly'' triggering is

\begin{equation}
p_{trig,incorrect}^{i,\left(2,delay\right)}=p_{trig\, back}^{i,\left(2,delay\right)}\left(1-p_{trig\, true}^{i,\left(2,delay\right)}\right).
\end{equation}

Then the total probability for one arm to trigger is

\[
p_{trig}^{i,\left(2,delay\right)}=p_{trig,correct}^{i,\left(2,delay\right)}+p_{trig,incorrect}^{i,\left(2,delay\right)}.
\]

When seeded with a pump laser with repetition rate $R$, the heralding
rate is then

\begin{equation}
R_{trig}^{i,(2,delay)}=R\times p_{trig}^{i,(2,delay)}.
\end{equation}

Following the techniques in \cite{bo-arx-1409-5341}, we can derive
the probability of single and multi photon detection given that the
heralding detector has \textit{not} triggered due to a paired idler
photon:

\begin{equation}
p_{single\, no\, trig}^{i,(2,delay)}=\frac{\left(1-\left|\xi\right|^{2}\left(1-\eta_{i}\right)\right)\eta_{s}\left(1-\eta_{i}\right)\left|\xi\right|^{2}}{\left(1-\left|\xi\right|^{2}\left(1-\eta_{i}\right)\left(1-\eta_{s}\right)\right)^{2}},
\end{equation}

\begin{equation}
p_{multi\, no\, trig}^{i,(2,delay)}=\left(1-\left|\xi\right|^{2}\left(1-\eta_{i}\right)\right)\left|\xi\right|^{2}\left(\frac{\left(1-\eta_{i}\right)}{1-\left|\xi\right|^{2}\left(\text{1-\ensuremath{\eta_{i}}}\right)}-\frac{\left(1-\eta_{s}\right)\left(1-\eta_{i}\right)}{1-\left|\xi\right|^{2}\left(1-\eta_{s}\right)\left(1-\eta_{i}\right)}\right)-p_{single\, no\, trig}^{i,(2,delay)}.
\end{equation}

Then the coincidence probability is
\begin{equation}
P_{C}^{\left(2,delay\right)}=p_{trig,correct}^{i,\left(2,delay\right)}\left(p_{single}^{i,(2,delay)}+p_{multi}^{i,(2,delay)}\right)+p_{trig,incorrect}^{i,\left(2,delay\right)}\left(p_{single\, no\, trig}^{i,(2,delay)}+p_{multi\, no\, trig}^{i,(2,delay)}\right).
\end{equation}

The expected coincidence rate is

\begin{equation}
R_{C}^{\left(2,delay\right)}=R\times P_{C}^{\left(2,delay\right)}.
\end{equation}

The accidental probability is

\begin{equation}
P_{A}^{\left(2,delay\right)}=p_{trig}^{i,\left(2,delay\right)}p_{trig}^{s,\left(2,delay\right)}.
\end{equation}

and the expected accidental rate is

\[
R{}_{A}^{\left(2,delay\right)}=R\times P_{A}^{\left(2,delay\right)}.
\]

The model which maximized the mean $R^{2}$ statistic for trigger counts,
coincidences, and accidentals was found for all delays. \foreignlanguage{english}{These
are shown in Table \ref{table2}.}

\begin{table}

\centering{}%
\begin{tabular}{|c|c|c|c|c|}
\hline 
Source  & $\eta_{i}$  & $\eta_{s}$  & $P_{seed}$ (mW)  & $\overline{R^{2}}$\tabularnewline
\hline 
Pass 2 Delay 0  & 0.018  & 0.0024  & 6.3  & 0.971\tabularnewline
\hline 
Pass 2 Delay 1  & 0.017  & 0.0021  & 6.7  & 0.982\tabularnewline
\hline 
Pass 2 Delay 2  & 0.016  & 0.0023  & 6.8  & 0.978\tabularnewline
\hline 
Pass 2 Delay 3  & 0.015  & 0.0020  & 6.9  & 0.987\tabularnewline
\hline 
\end{tabular}\protect\protect\caption{\textbf{Table showing source parameters for Pass 2.} Source parameters
were determined using a numerical optimization fit to the model. $\overline{R^{2}}$
is the mean $R^{2}$ from the triggering rate, coincidence, and accidental
fits.}
\label{table2}
\end{table}

\label{sec:singleenhance}

\subsection{Time Multiplexed Source Model: Pass 2}

The probability for the MUX source from Pass 2 to trigger is

\begin{equation}
p_{trig}^{MUX,2}=1-(1-p_{trig}^{i,(2,0)})(1-p_{trig}^{i,(2,1)})(1-p_{trig}^{i,(2,2)})(1-p_{trig}^{i,(2,3)}),
\end{equation}

and the expected heralding rate is

\begin{equation}
R_{trig}^{MUX,2}=R\times p_{trig}^{MUX,2}.
\end{equation}

The coincidence probability is

\begin{equation}
P_{C}^{MUX,2}=P_{C}^{\left(2,0\right)'}+\left(1-p_{trig}^{i,(2,0)}\right)\left(P_{C}^{\left(2,1\right)'}+\left(1-p_{trig}^{i,(2,1)}\right)\left(P_{C}^{\left(2,2\right)'}+\left(1-p_{trig}^{i,(2,2)}\right)P_{C}^{\left(2,3\right)'}\right)\right).
\end{equation}

where $P_{C}^{\left(2,delay\right)'}$ is given by $P_{C}^{\left(2,delay\right)}$
except with every instance of $\eta_{s}$ replaced with $\eta_{s}\eta_{sw}^{(2,delay)}$.

The expected coincidence rate is

\begin{equation}
R_{C}^{MUX,2}=R\times P_{C}^{MUX,2}.
\end{equation}

The accidental probability is

\begin{equation}
P_{A}^{MUX,2}=p_{trig}^{i,\left(2,0\right)}p_{trig}^{s',\left(2,0\right)}+\left(1-p_{trig}^{i,\left(2,0\right)}\right)\left(p_{trig}^{i,\left(2,1\right)}p_{trig}^{s',\left(2,1\right)}+\left(1-p_{trig}^{i,\left(2,1\right)}\right)\left(p_{trig}^{i,\left(2,1\right)}p_{trig}^{s',\left(2,1\right)}+\left(1-p_{trig}^{i,\left(2,2\right)}\right)p_{trig}^{i,\left(2,0\right)}p_{trig}^{s',\left(2,1\right)}\right)\right).
\end{equation}

where $p_{trig}^{s',\left(2,delay\right)}$ is given by $p_{trig}^{s,\left(2,delay\right)}$
except with every instance of $\eta_{s}$ replaced with $\eta_{s}\eta_{sw}^{(2,delay)}$.

The expected accidental rate is

\begin{equation}
R_{A}^{MUX,2}=R\times P_{A}^{MUX,2}.
\end{equation}

\subsection{Time and Space Multiplexed Source Model: Both Passes}

The probability for the complete multiplexed source to trigger is

\begin{equation}
p_{trig}^{MUX}=1-\left(1-p_{trig}^{MUX,1}\right)\left(1-p_{trig}^{MUX,2}\right),
\end{equation}

and the expected heralding rate is

\begin{equation}
R_{trig}^{MUX}=R\times p_{trig}^{MUX}.
\end{equation}

The coincidence probability is

\begin{equation}
P_{C}^{MUX}=P_{C}^{MUX,1}+\left(1-p_{trig}^{MUX,1}\right)P_{C}^{MUX,2}.
\end{equation}

The expected coincidence rate is

\begin{equation}
R_{C}^{MUX}=R\times P_{C}^{MUX}.
\end{equation}

The accidental probability is

\begin{equation}
P_{A}^{MUX}=P_{A}^{MUX,1}+\left(1-p_{trig}^{MUX,1}\right)P_{A}^{MUX,2}.
\end{equation}

The expected accidental rate is

\begin{equation}
R_{A}^{MUX}=R\times P_{A}^{MUX}.
\end{equation}

The expected CAR is

\begin{equation}
CAR^{MUX}=P_{C}^{MUX}/P_{A}^{MUX}.
\end{equation}

Using measured values of switching loss and electronics saturation,
the model was found to be in good agreement with the data (Main Text:
Fig. 4).

\section{Single-photon emission enhancement}

\begin{figure*}[th]
\includegraphics[scale=0.83]{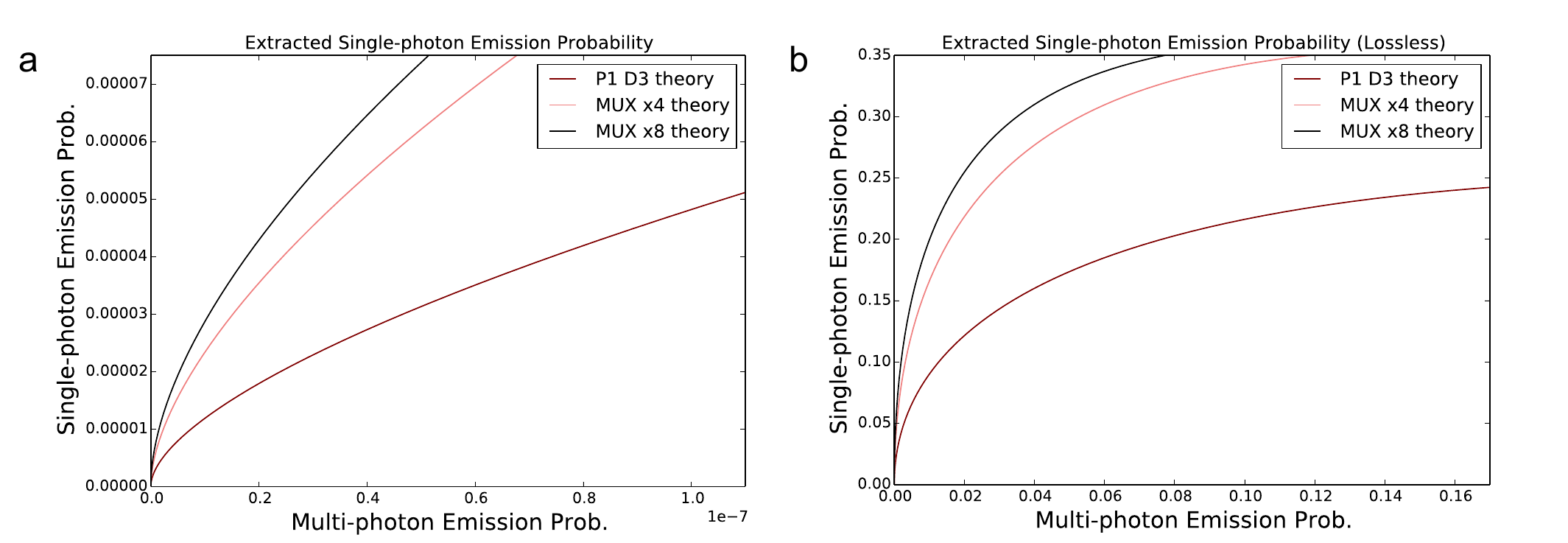}

\protect\protect\caption{\textbf{Extracted theoretical single-photon emission probability for
multiplexed and non-multiplexed sources. a,} Model with extrinsic
sources of loss removed. \textbf{b,} Model with all sources of loss
except switch loss and power loss removed.}

\label{Fig:ExtrTheoEmission}
\end{figure*}

Using the model with extrinsic sources of loss removed, we can extract
the heralded single-photon emission probabilities for fixed multi-photon
emission probabilities of the 8$\times$ and 4$\times$ multiplexed
sources and the best performing non-multiplexed source (Fig. S3a).
We include the loss of final measurement detector.

To further examine the long-term prospects of the multiplexed source,
we can also remove all sources of loss, except those due to the multiplexing
switches and power loss affecting the second pass. These sources of
loss include filtering, coupling, and detector inefficiencies (although
we still assume the use of threshold detectors). The extracted single-photon
emission probabilities for fixed multi-photon emission probabilities
of the 8$\times$ and 4$\times$ multiplexed sources and the best
non-multiplexed source are shown in Fig. S3b.

\end{document}